\documentclass[11pt]{article}
\usepackage{amssymb}
\usepackage{amsmath}
\usepackage{lscape}
\usepackage[dvips]{epsfig}
\def\nuc#1#2{\relax\ifmmode{}^{#1}{\protect\text{#2}}\else${}^{#1}$#2\fi}

\makeatletter
\newcommand{\l@vveden}[2]{\hbox to\textwidth{{\bf \quad #1 #2}}}

\makeatother

\begin{document}

{\bf \Large 
\noindent{
LANL Report LA-UR-00-3599, Los Alamos (2000)\\
}}

{\bf \Large 
\noindent{
Proc. SATIF5, July 18-21, 2000, Paris, France}}

\vspace*{1cm}

\begin{center}
{ \bf
THRESHOLD REACTION RATES INSIDE AND ON THE SURFACE OF 
THICK W-Na TARGET IRRADIATED WITH 0.8 GEV PROTONS} 


\vspace*{1.2cm}

{\bf      Yury E. Titarenko, Oleg V. Shvedov, Vyacheslav F. Batyaev, \\
Evgeny I. Karpikhin, Valery M. Zhivun, Aleksander B. Koldobsky, \\
Ruslan D. Mulambetov, Dmitry V. Fishchenko, Svetlana V. Kvasova }\\
 Institute for Theoretical and Experimental Physics, B. Cheremushkinskaya 25, Moscow, 117259, Russia

{\bf Andrey M. Voloschenko} \\
Keldysh Institute of Applied Mathematics, Miusskaya Sq. 4, 125047 Moscow, Russia

{\bf Stepan G. Mashnik, Richard E. Prael} \\
Los Alamos National Laboratory,  Los Alamos, NM 87545, USA

{\bf Hideshi Yasuda }\\
Japan Atomic Energy Research Institute, Tokai, Ibaraki, 319-1195, Japan

\vspace*{1.2cm}

\end{center}

\section*{Abstract}
\noindent

The preliminary results are presented of the experimental determination 
of threshold reaction 
rates in experimental samples made of Al, Co, Bi, In, Au,
to name but a few, placed both inside and on the surface of 
extended thick W-Na target irradiated with 0.8 GeV protons. 
The target consists of 26 alternating discs 
each 150 mm in diameter: 6 tungsten discs are 20 mm thick, 7 tungsten 
discs are 40 mm thick, and 13 
sodium discs are 40 mm thick. The relative position of discs is matched 
with the aim of flattening the 
neutron field along the target surface. The comparison is made of the 
measured rates with results of 
their simulation using the LAHET and KASKAD-S codes, and the ENDF/B6, 
MENDL2, MENDL2P, 
SADKO-2, and ABBN-93 databases. The results are of interest both in 
terms of integral data 
collection and to test the up-to-the-date predictive power of the
codes applied in designing of hybrid Accelerator Driven System
(ADS) systems that use tungsten targets cooled with sodium.

\section*{Foreword}
\noindent

Quantitative information on interaction of accelerated 
protons with different targets is 
necessary to design ADS. Application of hadron-nucleus 
process simulations should be 
tested by special experiments in which irradiation conditions, target 
material composition and location 
approximate the design type to the limit. There are several projects, 
\cite{bib1, bib2}, for example, where tungsten 
cooled by sodium is considered to be used as target material. 
This served as the basis for conducting 
experiments with micromodels of such targets. Comparison of experimental 
data obtained on such 
micromodels with corresponding calculated values will give us valuable 
information both for 
modifications of codes and databases and for assessment the 
calculation accuracy of the target part of 
the relevant ADS facility designs.

\section*{Experiment plan}
\noindent

The target with alternating adherent tungsten and sodium discs was designed 
in our micromodel. 
The location of these discs, which was specially chosen, 
facilitates the maximum flattening of the 
neutron field along the target. Experimental samples made of Al, Co, In, 
Au, Bi, \nuc{63}{Cu}, \nuc{65}{Cu}, \nuc{93}{Nb}, 
\nuc{64}{Zn}, \nuc{19}{F} (CF$_2$), \nuc{12}{C}, Ta, and Tm 
manufactured by punching the corresponding foils or by 
molding fine powders, 10.5 mm in diameter and 0.1--0.3 mm thick, 
were placed inside the target and on its surface. 

\begin{table}[t!]
\caption{Disc sequence, thickness, and experimental samples layout}
\begin{center}  \begin{tabular}{|c|c|c|c|c|}\hline \label{tab1}
Disc 	&	Material	&	Thickness, 	&	\multicolumn{2}{|c|}{Samples}	 \\ \cline{4-5}
number	&		&	mm	&	Inside	&	Outside	\\ \hline
W1	&	W	&	20	&	5Al*), Co	&	Al, In, Bi, Au, Ta	\\ \hline
Na1	&	Na	&	3*40=120	&	-	&	-	\\ \hline
W2	&	W	&	20	&	5Al, Co	&	Al, In, Bi, Au, \nuc{169}{Tm}, \nuc{63}{Cu}, Ta, 	\\ 
	&		&		&		&	\nuc{65}{Cu}, \nuc{93}{Nb}, \nuc{64}{Zn}, \nuc{19}{F}, \nuc{12}{C}, Co	\\ \hline
Na2	&	Na	&	2*40=80	&	-	&	-	\\ \hline
W3	&	W	&	20	&	5Al, Co	&	Al, In, Bi, Au, Ta	\\ \hline
Na3	&	Na	&	2*40=80	&	-	&	-	\\ \hline
W4	&	W	&	20	&	5Al, Co	&	Al, In, Bi, Au, Ta	\\ \hline
Na4	&	Na	&	40	&	-	&	-	\\ \hline
W5	&	W	&	40	&	5Al, Co	&	Al, In, Bi, Au, Ta	\\ \hline
Na5	&	Na	&	40	&	-	&	-	\\ \hline
W6	&	W	&	40	&	5Al, Co	&	Al, In, Bi, Au, Ta	\\ \hline
Na6	&	Na	&	40	&	-	&	-	\\ \hline
W7	&	W	&	40	&	5Al, Co	&	Al, In, Bi, Au, Ta	\\ \hline
Na7	&	Na	&	40	&	-	&	-	\\ \hline
W8	&	W	&	40	&	5Al, Co	&	Al, In, Bi, Au, Ta	\\ \hline
Na8	&	Na	&	40	&	-	&	-	\\ \hline
W9	&	W	&	40+20=60	&	5Al, Co	&	Al, In, Bi, Au, Ta	\\ \hline
Na9	&	Na	&	40	&	-	&	-	\\ \hline
W10	&	W	&	2*40=80	&	5Al, Co	&	Al, In, Bi, Au, Ta	\\ \hline
\multicolumn{2}{|c|}{}      &     W:  380     &         50 Al,  & 10 Al, 10 In, 10 Bi, 10 Au, \\
\multicolumn{2}{|c|}{ Total }&  Na: 520    &     10 Co &        10 Ta, 169Tm, 63¥u, 65Cu, 93Nb, \\
\multicolumn{2}{|c|}{}     &                     &             &     64Zn, 19F, 12C, Co   \\   \cline{3-5}
\multicolumn{2}{|c|}{}     &     900       &      \multicolumn{2}{|c|}{123 samples**)} \\ \hline 
  \end{tabular} \end{center} 

\vspace{0.1cm}
*) ``Al" designates a single Al sample; ``5Al" designates five Al 
samples, etc.

**)  There are samples (made of Al) beside those specified in Table 1,
 for measuring the proton beam density 
 distribution across the front (Al) surface and to control the neutron 
 field uniformity along discs that form the 
target.

\end{table}

\begin{figure} 
\centerline{\epsfxsize 14cm \epsffile{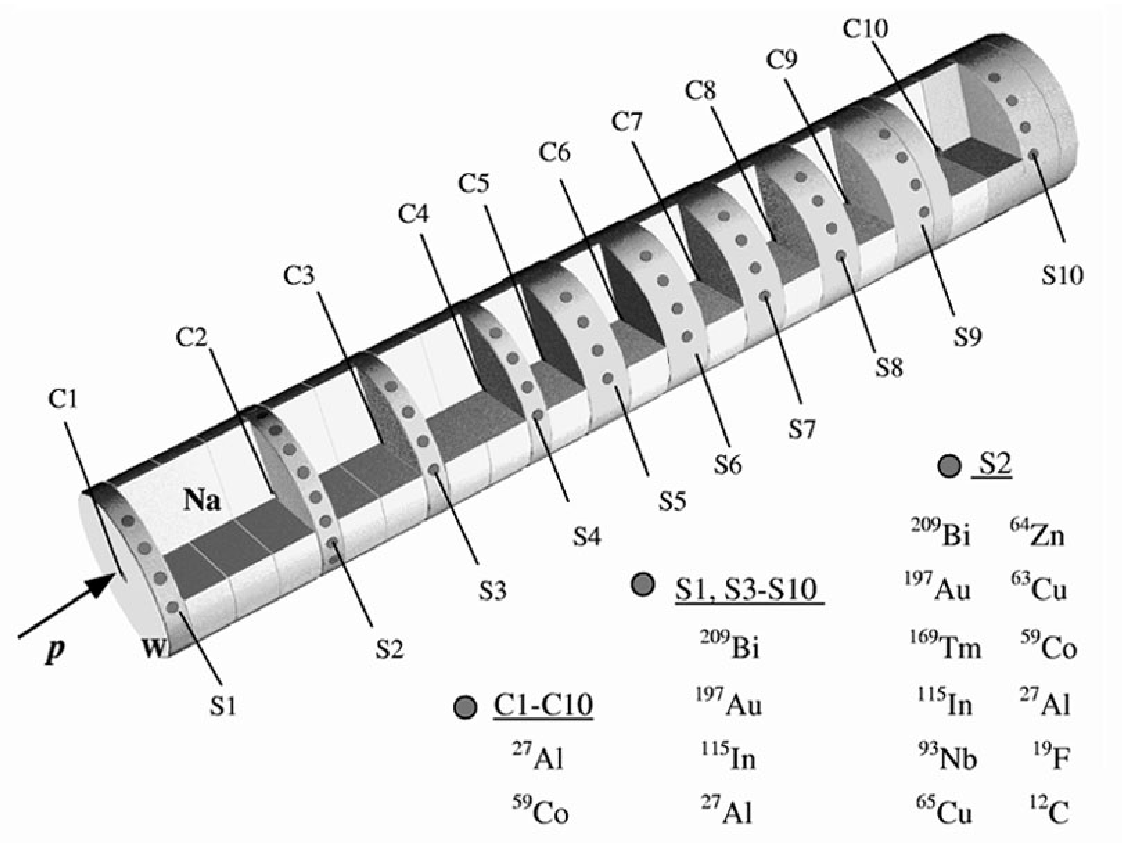}} 
\caption{Layout of W and Na discs and experimental samples.}\label{fig1} \end{figure}

The layout of tungsten and sodium discs is shown in 
Fig. \ref{fig1}. All discs are 150 mm in diameter, disc thickness 
and sequence is listed in Table \ref{tab1}. The tungsten 
discs have special design providing insertion of special bars with 
round recesses for experimental 
samples to be placed inside the target. The discs are located on a 
special adjustment table which 
provides alignment of target and proton beam axes with an accuracy of 
the order of 1 mm. The proton 
beam size examination has been performed by using in tentative exposures 
to radiation aluminum 
cut foils and Polaroid film.

W (97.5\%), Ni (1.75\%), Fe (0.75\%), and less than 0.2 \% of impurities 
are incorporated in 
tungsten discs. The average density of the tungsten discs is of 18.6 g/cm$^3$. 
Sodium discs represent metallic 
sodium placed into a steel container with 0.4 mm thick walls. 
Impurities content in Na is less than 
0.02\%.
 
\begin{figure} 
\centerline{\epsfxsize 12cm \epsffile{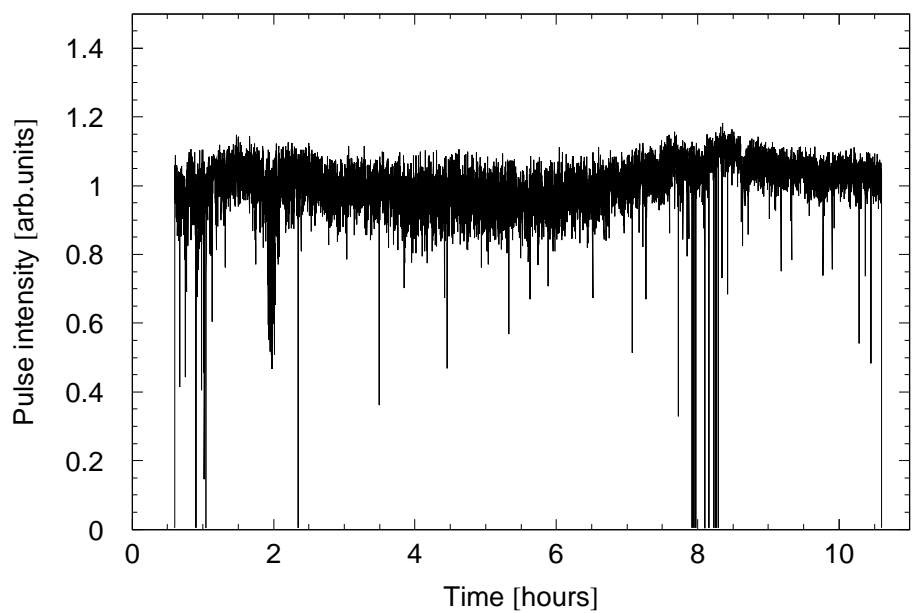}} 
\caption{Changes in the proton beam intensity over the period of irradiation.}\label{fig2} \end{figure}

Target irradiation was performed with 0.8 GeV protons over a period of 
10 hrs at an average 
intensity equal to 4.8$\cdot$10$^{10}$ p/cm$^2 \cdot$pulse using the ITEP 
synchrotron. Pulse repetition rate is of 15 pulses per 
minute. Changes in the proton beam intensity over the irradiation period is 
presented in Fig. \ref{fig2}. After a short 
decay lag, experimental samples were extracted from the target's 
surface and inside volume and 
packaged into labeled polyethylene packages. Subsequent gamma spectra 
measurements were made  
using several spectrometers. The absolute value for different threshold 
reaction rates were determined 
using the PCNUDAT decay database after gamma-spectra processing 
with the GENIE2000 code (see Table \ref{tab2}).
 
\section*{Simulation of reaction rates}
\noindent

To simulate the measured reaction rates,
 the LAHET Code System (LCS) \cite{bib3} was used which 
involves the LAHET code for simulating hadron-nuclei interactions, 
the HMCNP code for simulating 
neutron transport at energies below 20 MeV, and the 
PHT code for simulating hard photon transport. 
Both neutron and proton spectra were calculated in all points where 
experimental 
samples are located. 

Another code system applied to simulate the experiment is the 
KASKAD-S code \cite{bib4} which 
uses a discrete ordinate algorithm for coupled charges/neutral particle 
transport calculations in 2D-pencil beam problems. 
The multigroup cross-section library SADKO-2 for nucleon-meson cascade 
calculations coupled with the CONSYST/ABBN-93 neutron and 
gamma-ray cross-section libraries below 20 MeV is used.

\begin{table}[t!]
\caption{Thresholds and main parameters of products for 
reaction rates being measured}
 \begin{center} \begin{tabular}{|c|c|c|c|c|}\hline \label{tab2}
Reaction	&	E$_{th}$(MeV)	&	Product	&	T$_{1/2}$	&	E$_{\gamma}$(keV) (Y$_{\gamma}$(\%))	\\ \hline
\nuc{115}{In}(n,n')	&	0.335	&	\nuc{115m}{In}	&	4.49h	&	336.2(45.8)	\\ \hline
\nuc{27}{Al}(n,p)	&	1.89	&	\nuc{27}{Mg}	&	9.46m	&	843.7(71.8)	\\ \hline
\nuc{64}{Zn}(n,p)	&	2.1	&	\nuc{64}{Cu}	&	12.70h	&	511.00(35.8)	\\ \hline
\nuc{59}{Co}(n,p)	&	2.6	&	\nuc{59}{Fe}	&	44.50d	&	1099.3(56.5)	\\ \hline
\nuc{27}{Al}(n,$\alpha$)&	3.26	&	\nuc{24}{Na}	&	15.02h	&	1368.5(100)	\\ \hline
\nuc{197}{Au}(n,2n)	&	8.5	&	\nuc{196}{Au}	&	6.18d	&	355.7(86.9)	\\ \hline
\nuc{115}{In}(n,p)	&	9	&	\nuc{115}{Cd}	&	2.23d	&	527.9(27.5)	\\ \hline
\nuc{115}In(n,2n)	&	9	&	\nuc{114}{In}	&	49.51d	&	191.6(16)	\\ \hline
\nuc{65}{Cu}(n,2n)	&	10.1	&	\nuc{64}{Cu}	&	12.70h	&	511.00(35.8)	\\ \hline
\nuc{63}{Cu}(n,3n)	&	$\sim$15	&	\nuc{61}{Cu}	&	3.408h	&	656.0(10.7)	\\ \hline
\nuc{59}{Co}(n,2n)	&	11	&	\nuc{58}{Co}	&	70.92d	&	810.8(99.4)	\\ \hline
\nuc{19}{F}(n,2n)	&	11.1	&	\nuc{18}{F}	&	1.83h	&	511(194)	\\ \hline
\nuc{64}{Zn}(n,2n)	&	12.4	&	\nuc{63}{Zn}	&	38.1m	&	669.6(8.40)	\\ \hline
\nuc{12}{C}(n,2n)	&	$\sim$20	&	\nuc{11}{C}	&	20.38m	&	511(162)	\\ \hline
\nuc{209}{Bi}(n,4n)	&	22.6	&	\nuc{206}{Bi}	&	6.24d	&	803.1(98.9)	\\ \hline
\nuc{197}{Au}(n,4n)	&	23	&	\nuc{194}{Au}	&	39.5m	&	328.4(63)	\\ \hline
\nuc{169}{Tm}(n,4n)	&	25.5	&	\nuc{166}{Tm}	&	7.7h	&	778.82(19.9)	\\ \hline
\nuc{209}{Bi}(n,5n)	&	29.6	&	\nuc{205}{Bi}	&	15.31d	&	1764.3(32.5)	\\ \hline
\nuc{93}{Nb}(n,4n)	&	31	&	\nuc{90}{Nb}	&	14.6h	&	1129.2(92.7)	\\ \hline
\nuc{115}{In}(n,5n)	&	35	&	\nuc{111}{In}	&	2.83d	&	245.4(94)	\\ \hline
\nuc{209}{Bi}(n,6n)	&	38	&	\nuc{204}{Bi}	&	11.22h	&	899.2(98.5)	\\ \hline
\nuc{209}{Bi}(n,7n)	&	45.3	&	\nuc{203}{Bi}	&	11.76h	&	820.2(29.6)	\\ \hline
 \end{tabular} 
{}
\end{center}
\end{table}

Reaction cross sections for neutrons with energies up to 100 MeV and 
protons with energies up to 
200 MeV were taken from MENDL2 \cite{bib5} and MENDL2P \cite{bib6} 
libraries, respectively. 
Reaction 
rates
are obtained via integral product of spectra and cross sections
 ($R=\int\phi(E)\sigma(E)dE$).
The discrepancy between calculated and experimental results was estimated 
using
the root-mean-square discrepancy factor,
$<$F$>$, 
defined in \cite{bib7}.

\section*{Results}
\noindent

\begin{figure}[t!] 
\centerline{\epsfxsize 11cm \epsffile{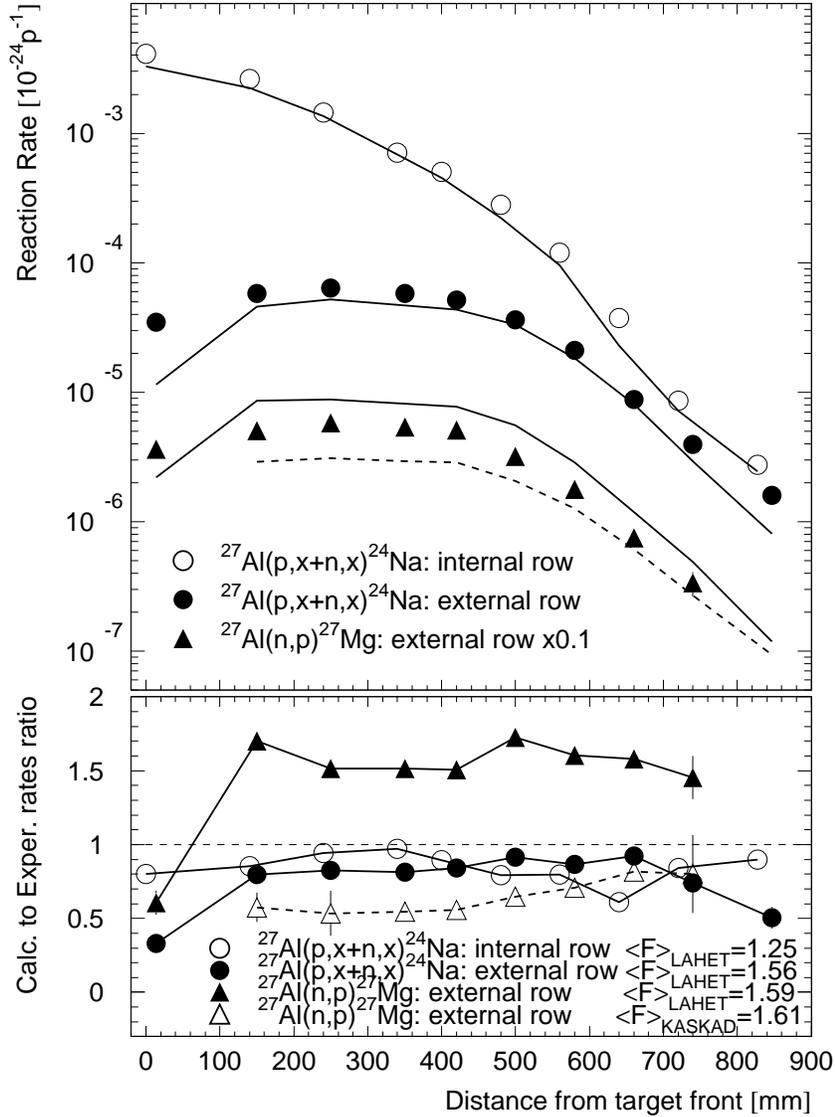}} 
\caption{Reaction rates of \nuc{27}{Al}(p,x+n,x)\nuc{24}{Na} and 
\nuc{27}{Al}(n,p)\nuc{27}{Mg}
 at the target axis and on the surface (top plot). 
 Calculations by 
LAHET and KASKAD-S are shown by solid and dashed lines, 
respectively. The ratios of reaction rate calculated values to 
the measured data are shown on the bottom plot.}\label{fig3} \end{figure}

The preliminary processing of gamma spectra from Al samples 
placed on the target outside 
surface and in the target center makes it possible to estimate the
generation rates of \nuc{24}{Na} and \nuc{27}{Mg}. 
Experimental and calculated values obtained using the LAHET code 
are presented in Fig. \ref{fig3}. 
Measurements of gamma-spectra for all the samples irradiated are 
still being in progress and all the 
reaction rates listed in Table \ref{tab2} are expected to be 
determined after the completion of gamma-spectra processing.

\section*{Convergence of calculated and experimental values}
\noindent

The rates for \nuc{24}{Na} in Al foils predicted by LAHET 
are in satisfactory agreement with 
experimental values (except the first and last points
 on the surface), as shown in Fig. \ref{fig3}. 
Whereas,
the predicted rates for \nuc{27}{Mg} generation on the target surface 
are systematically 
overestimated 
 compared to the measured values 
(by a factor of 1.6, on average), 
except the first point.
 At the same time, the
\nuc{27}{Mg} production rates predicted by KASKAD-S  are underestimated
on the average by a 
factor of 1.6. 
The observed
discrepancies would be better understood
only after the complete 
processing of gamma spectra for the rest of samples and determining 
both (n,p) --  purely ``neutron"   
and (p,n) -- purely  ``proton"  reactions.

\section*{Acknowledgement}
\noindent

The work was carried out under the
ISTC Project\#1145 and was supported by the JAERI (Japan) and, 
partially, by the U. S. Department of Energy.

\end{document}